\documentstyle[multicol,pre,aps,epsf,psfig]{revtex}
\draft
\begin{document}
\newcommand{\bea}{\begin{equation}}
\newcommand{\ber}{\begin{eqnarray}}
\newcommand{\eea}{\end{equation}}
\newcommand{\eer}{\end{eqnarray}}
\newcommand{\ct}{\cite}
\newcommand{\bi}{\bibitem}

\title{The role of pinning and instability in a class of non-equilibrium growth models}
\author{ Amit K. Chattopadhyay}
\address{
Max Planck Institute for the Physics of Complex Systems, 
N\"othnitzer Strasse 38, 
D-01187 Dresden, Germany}
\date{\today}
\maketitle

\footnote{email: akc@mpipks-dresden.mpg.de}
\begin{abstract}
We study the dynamics of a growing crystalline facet where the growth mechanism is controlled by the geometry of the local
curvature. A continuum model, in (2+1) dimensions, is developed in analogy 
with the Kardar-Parisi-Zhang (KPZ) model is considered for the purpose.
Following standard coarse graining procedures, it is shown that in the large time, long distance limit, the continuum model
predicts a curvature independent KPZ phase, thereby suppressing all explicit effects of curvature and local pinning in the
system, in the "perturbative" limit. A direct numerical integration of this growth equation, in 1+1 dimensions,
supports this observation below a critical parametric range, above which generic instabilities, in the form of isolated
pillared structures lead to deviations from standard scaling behavior. Possibilities of controlling this instability by
introducing statistically "irrelevant" (in the sense of renormalization groups) higher ordered nonlinearities have also been
discussed. 
\end{abstract}
\begin{multicols}{2}
\section{Introduction}
Growth of interfaces in a strongly temperature controlled 
regime has recently been a subject of considerable 
technological importance [1]. The interest in this field has mainly been generated by the microscopic
roughness that originates as a result of competition among different effects,
such as surface tension, thermal diffusion and different noise factors coming
into play during the growth process. Even in equilibrium, a crystalline facet
"remains practically flat until a transition temperature is reached, at which
the roughness of the surface increases very rapidly" [2]. Numerous theoretical
models, starting from Burton, et al [2] have been proposed to account for a
detailed analysis of the height fluctuations during the growth process on
both sides of the roughening temperature $ T_R $ [3-5]. They found a change
in mobility from activated growth in the nucleated phase (characterized by
low temperatures) to non-activated growth at large temperatures. The
pinning-depinning growth model of Chui and Weeks [3] has later been
numerically extended to a polynuclear growth model by Sarloos and Gilmer [6]
confirming a growth by island formation which is continuously destroyed by
any small chemical potential favoring the growth. 
On a different variety of non-equilibrium growth, phase boundaries of growing fingers emerging out of a competition between morphological instabilities with the stabilizing effect of surface tension have also been studied to get an idea of how systems arrive at their observed patterns [7]. Theoretical forays in this
front have continued with detailed predictions on the equilibrium roughening
transition [8,9] which have also found experimental justifications in
[10]. The theoretical efforts [8,9] centered around driven sine-Gordon
models, coupled with relevant non-linearities and then looking at
possible roughening transitions. The experimental value [10] of the
roughening temperature compared extremely favorably with the theories.
\par
Starting from the early theoretical efforts [2,3] to the present day
developments [1,11], numerous discrete [12,13] and continuum [14-16] models
have been proposed to study both equilibrium and non-equilibrium surface
properties. Historically speaking, most of these nonlinear models seem to fall 
within either of the KPZ or the Lai-Das
Sarma (LD) ({\it i. e.} conserved and non-conserved) universality class, barring a few exceptions [12,16,17]. The atomistic
growth process of the latter type deals with surfaces grown by molecular beam
epitaxy (MBE) method, whose distinguishing feature is that there the growth occurs
under surface diffusion conditions, with the deposited atoms relaxing to the
nearby kinks. The essential idea employed is to modify the relaxation
mechanism as a locally surface minimizing curvature.
To linear order, this mechanism is supposed to mimic the growth dynamics of
crystalline surfaces [13,15].

However, the basic atomistic feature associated with either of these universality classes concerns a growth restriction on
the height difference between competing sites. Only relatively few digressions are known, of which 
a classic example is the equilibrium restricted-curvature (RC) model studied by Kim and Das Sarma [18]. The corresponding
growth rule describes a growth restricted on the local curvature,
$ \mid{{\nabla}^2 h}\mid \leq N $, and is obeyed at both the growing site
and its nearest neighbors where $ N $ is any fixed positive integer. Particles are not allowed to stick to a particular site if this growth restriction is not met with. We adopt
the non-equilibrium version of the growth model proposed by Kim and Das Sarma as the
starting point of our study of the dynamics of the non-equilibrium growth in an
MBE process. Our basic theoretical approach follows the standard dynamic renormalization scheme employed as in [8,9]. The
technical details are elaborated in section 3 and a brief discussion on the implication of these results, from the standpoint
of non-equilibrium critical phenomena, is provided in the following section.

The remaining portion of the paper is a numerical investigation of the same problem, starting from the same dynamic equation
as used in the analytical approach. We attempt a direct numerical integration of this dynamical equation using Eulerian
discretization schemes. However, here we had to restrict ourselves in 1+1 dimensions due to the insurgence of certain generic
instabilities associated with the growth process. Here it must be stressed that as observed in [19] and something which
has been strongly supported by our numerical observations too, the instabilities that we see, have got nothing to do with the
discretization scheme but is, in fact, a characteristic feature of the growth process itself. Identical observation,
in context to different models have also been reported in [20,21]. We have checked with both the deterministic and stochastic
versions of our model and have found more or less identical results as in [19], although, in our case, contributions from a
periodic pinning potential do show up in certain parametric limits. However, 
most significantly, a continuum analysis of this
same model never props up with any such atomistic singularities, a result, 
which again is not unlike to those found in [19].
The reasons for this is not at all difficult to understand, since, in any continuum approach, we are essentially building
upon a perturbative analysis where the tacit assumption is always a vanishingly small nearest-neighbor curvature difference
(our case) or an infinitesimally small height difference (previous studies) between neighboring sites. But the catch lies in
the fact that the algebra of derivatives cannot be analytically continued to the domain of difference operators, if the
nearest-neighbor curvature difference (or heights, depending on the definition of the model) is non-negligible, as was
rightfully observed by Dasgupta, {\it et al}. This automatically implies that 
any nonlinear discretized process might be substantially
different from its continuum analogue, even though it is derived from
the same Langevin equation. However, for linear growth models, there is no 
such disagreement between the discrete and continuum theories.

We study all these effects in some detail and the results are explicitly described in section 5. In the following section, we
discuss the technical possibilities of controlling these instabilities (meaning sudden growth of a hillock or a hump in an
arbitrary site, as compared to its neighbors) in the same line as in [19-21]. In this respect, our emphasis is on the method
of introducing "irrelevant" (in the sense of renormalization) higher ordered nonlinear operators to muffle these generic
instabilities while still retaining the structure of the basic universality class unperturbed. Our numerical results in 1+1 dimensions seem to
clearly indicate that the model we are considering belongs to the Kardar-Parisi-Zhang (KPZ) [14] universality class in the
perturbative limit of the periodic potential. However, in the "strong coupling" limit of this periodic pinning potential, higher
harmonics of periodicity seem to hold sway and then our perturbative analysis fails, although we still seem to get 
numerically stable values for exponents, which certainly are quite different from the KPZ ones. At the present moment, we
do not have any analytical technique to study this "strong-coupling limit" (in the sense of a strong periodic potential) but
our achievement lies in the development of a stable numerical scheme. Finally we conclude with a
comparative discussion of numerical and analytical results in section 7.

\section{The model and its variations}

Considering two dimensional growth pertaining to a lattice structure where
$ h(\vec r,t) $ is the height of the interface at time $ t $ at position
$ \vec r $, the equation goes like

\ber
\eta \frac{\partial h}{\partial t} &=& -\gamma {\nabla}^4 h(\vec r,t) +
\frac{\lambda}{2} {\mid {\nabla}^2 h(\vec r,t) \mid}^2  \nonumber \\
&-& \frac{2\pi V}{a}
\sin[\frac{2\pi h(\vec r,t)}{a}] + F + R(\vec r,t)
\eer

{\noindent}
where $ \eta $ is the inverse mobility which fixes the time scale, $ \gamma $
is the surface tension, $ a $ is the lattice constant and $ V $ is the
strength of the pinning potential. $ F $ is a steady driving force with 
the white noise R being defined as

\bea
< R(\vec r,t) R(\vec r',t') > = 2D {\delta}^2(\vec r-\vec r') \delta(t-t')
\eea

{\noindent} From a fluctuation-dissipation theorem, in a static
version of the model, we relate the noise-strength D to the equilibrium temperature, by the
relation $ D = \eta T $ [9,25]. The biharmonic term on the right hand side of the above equation gives the basic
relaxation mechanism in operation. The second term is the most important term
for the restricted-curvature (RC) dynamics and has been incorporated in
analogy with the KPZ equation [14]. Just as the lateral term $ {\mid{\vec
\nabla h}\mid}^2 $ gives the nonlinearity in a KPZ growth, the curvature
dependence in our model is proposed to produce a leading order nonlinearity
of the form $ ({\nabla}^2 h)^2 $. From phenomenological considerations,
since the discretisation  scheme prohibits a high curvature, the constant
$ \lambda $ is negative. Although this explains the convex hillocks appearing on the growing surface, the restraint on the 
unbounded growth of concave structures actually come from the appearance of a linear curvature term in eqn.(7) from a 
renormalization of the sine-Gordon potential. This is basically the reason for which we do not need to start with a linear 
curvature term in the starting dynamical equation, unlike as in [22,23], rather it is automatically generated as a product of the dynamics of the 
growth process [16]. Actually we can do this since we are not
considering any step-edge barriers. With identical logic, since, to
start with, there is no boundary condition on the lateral growth property of
the surface, a KPZ-type non-linearity is not required in the system,
although this is indeed produced after a first renormalization of the
sine term. Apparently one might think that this makes it an "open" system
in a field-theoretic sense, but the point of concern for us is the
phenomenology of the growth process, where we are only concerned with
a curvature restriction in a periodic profile, completely
neglecting possibilities of other initial conditions. What the $ {({\nabla}^2 h)}^2 $ term does, therefore, is to provide the first order non-linear fluctuations 
around the steadily growing front in the RC phase. The third term comes along due to the lattice
structures  preferring integral multiples of $ h $, which means that $ h $ is
measured in units of the lattice constant . As to the sign of V, we see that changing $ V \rightarrow -V $, as $ h \rightarrow -h $, 
equation (1) remains unchanged in the absence of the curvature term. But in the presence of the curvature term, a negative V,
simply means that the periodic potential will decay to zero in the long time limit and the system will tend towards the 
Govind-Guo universality class [27]. So, for all practical purposes, V is positive. $ F $ is the steady driving force required to
depin a pinned interface while all the microscopic fluctuations in the growth
process are assimilated in the noise term $ R $. It can be shown easily that
the second and third terms are obtained from a variant of the stationary-state sine-Gordon
Hamiltonian

\bea
E[h(\vec r,t)] = \int\int d^2r [\frac{\gamma}{2} ({\nabla}^2 h)^2 -
V \cos(\frac{2\pi}{a} h)]
\eea

{\noindent}
In the following analysis of the dynamic model, we will try to probe the existence of a roughening transition, akin to that
in [8,9] and have an analytical estimate of this transition temperature.
This obviously brings into question the interplay of
strengths between the non-linear term $ ({\nabla}^2 h)^2 $ and the sine-Gordon
potential. Three subsets of our model, defined through eqn.(1), 
are well known: \\
(i) $ \lambda = V = F = 0 $; characterizing a stationary interface
[11,12,17]. For example, in the Das Sarma-Tamborenea model [12], it was
found that the growth dynamics was intermediate to that between the
random-deposition model with no relaxation (RD) and the
random-deposition model with perfect relaxation (RDR). The growth
exponent $ \beta (\sim 0.375) $ in this model is intermediate between that 
of the RD ($\beta \sim  0.5 $) and the RDR ($\beta \sim  0.25 $) models.  
This sort of growth has found experimental justification in the works of Yang,
et al and Jeffries, et al [23]. \\
(ii) $ \lambda = F = 0,\:\:V \neq 0 $; characterizing the growth dynamics of
crystalline tensionless surfaces [25]. This model in equilibrium depicts a
roughening transition to the high temperature regime of the sine-Gordon model
and is expected to model the vacuum vapor deposition dynamics by MBE growth. \\
(iii) $ V = F = 0,\:\:\lambda \neq 0 $; a very special case of the numerical
simulation predicting a "local model" for dendritic growth with
conventional renormalization
failing to define a stable, non-trivial fixed point [27]. However
numerically it is possible to determine the critical exponents as has
been done bby Govind and Guo too. The interesting open problem here is
how to arrive at these values analytically.
\par
However in both the first and second situations, detailed numerical analysis
have predicted a transition from the conserved MBE growth to the Edwards-Wilkinson [28]
type phase characterized by the generation of a
$ {\nabla}^2 h $ term [25,29]. The generation of this so-called "surface
tension" term in the dynamics although surely being a fall-out of the
renormalization of the sine-Gordon potential in the latter model, actually
demands a greater attention. Combining this idea with the proposition put
forward by Kim and Das Sarma [18], we therefore pose the most general
situation concerning a surface growing under MBE and try to ascertain the
associated roughening process throughout the whole temperature range, with
the growth essentially occurring under a surface curvature constraint. 

\section{Dynamic renormalization and the results obtained}

In the following analysis, we employ standard dynamic renormalization techniques to
probe the dynamics of our proposed Langevin equation (1). 

We start by first integrating over the momentum shell
$ \Lambda(1-dl) < \mid \vec k \mid < \Lambda $ perturbatively in $ \lambda $
and $ V $. Thereafter we rescale the variables back in the form
$ \vec k \rightarrow \vec k' = (1+dl)\vec k,\:\:h \rightarrow h' = h $ and
$ t \rightarrow t'=(1-4dl)t $. The various coefficients follow the rescaling
$ \eta \rightarrow \eta' = \eta,\:\:\gamma \rightarrow \gamma' = \gamma,
\:\:\lambda \rightarrow \lambda' = \lambda,\:\:V \rightarrow V' = (1+4dl)V,
\:\:F \rightarrow F' = (1+4dl)F $. We set up the perturbative scheme to
rewrite eqn.(2) in a comoving frame of reference moving with velocity $ F/\eta $ as

\bea
\eta \frac{\partial}{\partial t} h = -\gamma {\nabla}^4 h + \Phi(h) + R
\eea

{\noindent} where $ \Phi(h) = -\frac{2\pi V}{a}\:\sin[\frac{2\pi}{a}(h+\frac{F}{\eta} t)]
+ \frac{\lambda}{2} ({\nabla}^2 h)^2 $.
Now, following the line of Nozieres and Gallet [8[ and Rost and
Spohn [9], we divide both h($\vec x$,t) and R as sums of two parts, one an 
average part $\bar h$, $ \bar R $ and the other a fluctuating part 
$ \delta h $, $ \delta R $.
This latter fluctuating part actually defines the short-wavelength components of noise and is defined within the
momentum shell $ \Lambda (1-dl) < |\vec k| < \Lambda $. 
This gives

\ber
\bar \Phi &=& -\frac{2\pi}{a}\:V \sin[\frac{2\pi}{a}(\bar h + \frac{Ft}{\eta})]\:
(1-\frac{2{\pi}^2}{a^2} <\delta h^2>) \nonumber \\
&+& \frac{\lambda}{2}{({\nabla}^2 {\bar h})}^2 + \frac{\lambda}{2} 
<{({\nabla}^2 {\delta h})}^2>
\eer

and

\bea
\delta \Phi = -\frac{4{\pi}^2}{a^2} V \cos[\frac{2\pi}{a}(\bar h + \frac{F t}{\eta})] \delta h + \lambda {\nabla}^2{\delta h}.{\nabla}^2{\bar h} 
\eea

Now averaging out the short-wavelength components
and including the corrections (as in [30]), we finally arrive at an
expression for the renormalized mode coupling term,

\ber
\Phi^{SG} &=& -\frac{2{\pi}^3 V^2 T}{{\gamma}^2 a^5} dl\:\int_{-\infty}^t\:
dt' \int d^2r'\:\frac{1}{t-t'}\: J_0(\Lambda(\mid \vec r-\vec r'\mid) \nonumber \\
&\times&
G_0(\mid \vec r-\vec r' \mid, t-t') \nonumber \\ 
&\times& e^{-[\frac{\gamma}{\eta}{\Lambda}^4(t-t') + \frac{2\pi T}{a^2
\gamma} \phi(\mid \vec r-\vec r' \mid,t-t')]} \nonumber \\
&\times& [\frac{2\pi}{a}[\frac{\partial}{\partial t} \bar h(\vec r,t) (t-t')
\nonumber \\
&-& \frac{1}{2} {\partial}_i {\partial}_j \bar h(\vec r,t) (r_i-r_i')
(r_j-r_j')]\:\cos[\frac{2\pi}{a} \frac{F}{\eta} (t-t')] \nonumber \\
&+& [1-\frac{2{\pi}^2}{a^2} [{\partial}_i \bar h(\vec r,t)]^2 (r_i-r_i')^2]
\:\sin [\frac{2\pi}{a} \frac{F}{\eta} (t-t')]]
\eer

{\noindent} where $ \Lambda \sim 1/a $ is a suitably chosen upper cut-off with the
Green's function $ G(x,t) $ given by

\bea
G({\vec x},t)= \int d{\vec k}\:e^{i{\vec k}.{\vec x}-\frac{\gamma}{\eta} t k^4} 
\eea

and 

\bea
\phi(\tilde \rho,x) = \int_{\Lambda_0}^{\Lambda}\:\frac{dk}{k}\:[1-J_0(k{\tilde \rho})]\:
e^{-\frac{\gamma}{\eta} k^4 (t-t')}
\eea

{\noindent} with $ \tilde \rho = \Lambda \rho $, $ x = \frac{\gamma(t-t')}
{\eta {\rho}^2} $ and $ \Lambda_0 \sim 1/L $, is the inverse of the system size.
\par
Turning now to the above eqn.(7), we find that starting with a structurally
non-linear term in the Langevin equation, the dynamic renormalization has
initiated a mode coupling structure with quite a different composition 
compared to [8,9].
The absence of the nonlinear $ ({\nabla}^2 h)^2 $ term in $ {\Phi}^{SG} $
implies that once starting with a restricted curvature model of growth, the
lattice structure partakes a dynamics, where starting from an initial curvature dominated phase, $ \lambda $ is 
later renormalized to zero and then at large time limit, the KPZ type 
nonlinearity takes over. A relative comparison of the coefficients of the RC 
term, as given in eqn.(1) and from the 
renormalized KPZ term, obtained from eqn.(7), easily shows that the ratio of 
their respective strengths is equal to $ \frac{\lambda}{\lambda'} a^2 $, 
where $ a $ is the lattice constant and $ \lambda $ and $ \lambda' $ are the
strengths of the RC and KPZ nonlinearities respectively. To begin with, at 
very small times, since a 
steady growth occurs in the phase, the RC term dominates. But shortly, the 
pinning structure takes over the dynamics  
and from there onwards, the KPZ term comes into the picture. Thus, although a competition ensues between the alternate
pinning and depinning forces offered by the $ (\nabla h)^2 $ and
$ ({\nabla}^2 h)^2 $ nonlinearities, at short time scales, for all practical purposes, in the thermodynamic limit, the
periodic nature of the potential governs and the resultant scaling
structure shows a KPZ type behavior. In fact, the reason for which
conventional RG works in our model, whereas an apparently simpler
version of it, the Govind-Guo model, comes up with an unstable fixed
point, is due to the cross-over of our model to the KPZ type, in the
long time, large length scale limit. In the following
section, we tackle the problem numerically, which reconfirms our analytical 
prediction. The average $ <\delta h^2> $ renormalizes $ V $ and correction to 
the one-loop order looks like

\bea
dV^{(1)} = -\frac{2{\pi}^2}{a^2}\:<{(\delta h^{(0)})}^2> = -\frac{\pi T}{a^2 
\gamma} V dl
\eea

{\noindent} where $ V^{(1)} $ is the first-ordered perturbation in potential V
and $ \delta h^{(0)} $ is the free height fluctuation term. 
From an analysis of the above equation, we arrive at the following renormalized flow equation which defines the cross-over temperature:

\bea
\frac{dU}{dl} = (4-n)U
\eea

{\noindent} where $ U=\frac{V}{{\Lambda}^2} $ and $ \mathrm {n}=\frac{\pi T}
{\gamma a^2} $. The above equation indeed is a very important one and 
provides a quantitative description of the effects of curvature restriction in 
the dynamics of our model. However, since the other flow equations do not 
really have any fundamental bearing on our discussion, we remove them to the
appendix section for the more technically minded reader.
Eqn.(8) clearly shows that in the limit $ n = \frac{\pi T}{\gamma a^2} > 4 $, U renormalizes to zero which means that in the
large scale, the effects of the lattice potential vanish and the KPZ phase takes over. As consistency checks, we easily see
that putting $ \gamma = \lambda = 0 $, we reproduce the results of Rost and Spohn [9] and fixing $ \gamma = \lambda =
\lambda' = 0 $, the Nozieres-Gallet [8] results are obtained. The roughening transition occurs at $ T_R = \frac{4\gamma
a^2}{\pi} $, in suitable units. This sort of a roughening transition was also observed in an earlier work by Tang and Nattermann [16].
From this point, qualitatively, the whole dynamics follows in the line 
predicted by Nozieres and Gallet and
Rost and Spohn, with the biharmonic and structural nonlinear terms
muffled by the surface tension term and the KPZ nonlinearity respectively.
This KPZ regime continues until the growing interface reaches the next
crystalline facet whereon the RC dynamics again comes into play and the whole
mechanism goes on repeating itself throughout the process of non-equilibrium
growth. Although this apparently would imply the "irrelevance" of the 
restricted-curvature term, but the effect of this curvature restriction is
impregnated in the value of the critical temperature which is twice the value
obtained in [8,9], an outcome of the fact that n=4 in this case, as compared
to n=2 in [8,9]. Thus the information of the local geometry is conveyed through
the critical temperature, and although one finds a proliferation of interaction
 as in the present case, the RG needs not be redone. As an added information, we might add that this value of the critical temperature is not obvious from an analysis of the scaling dimenion of the system and one needs RG to arrive at this exact value.   

\section{Instability in the discrete growth process}

In this section we will describe in detail results obtained from a direct numerical integration of the restricted-curvature (RC)
model, in 1+1 dimensions, and the related instabilities in the growth process itself. The discretization process which we employ
here is a standard finite-difference scheme, defined in the following manner: \\

\bea
{\nabla}^2 f(\vec r,t) = f(\vec r + \triangle {\vec r},t) + f(\vec r- \triangle {\vec r},t) - 2f(\vec r,t)
\eea

{\noindent} and

\ber
{\nabla}^4 f(\vec r,t) &=& {\nabla}^2 ({\nabla}^2 f(\vec r,t) \nonumber \\
&=& {\nabla}^2 \lbrack{f(\vec r + \triangle {\vec r},t) + f(\vec r- \triangle
{\vec r},t) - 2f(\vec r,t)}\rbrack \nonumber \\
&=& f(\vec r+2\triangle {\vec r},t) + f(\vec r-2\triangle {\vec r},t) \nonumber \\
&-&
4\:f(\vec r + \triangle {\vec r},t)
- 4\:f(\vec r- \triangle {\vec r},t) + 6\:f(\vec r,t)
\eer

{\noindent} where all the variables h, $ \vec {\mathrm {r}} $, t have been suitably non-dimensionalised. In solving the above equations numerically, we start from a flat surface and invoke periodic boundary conditions. 

The quantities of our interest are the spatial and temporal correlation functions, defined as follows: \\
spatial correlation function $ \equiv $ $ {\langle {| h(\vec r+\vec r',t) - h(\vec r',t) |}^2 \rangle}_{\vec r'} $, \\
temporal correlation function $ \equiv $ $ {\langle {| h(\vec r,t+t')
    - h(\vec r,t') |}^2 \rangle}_{t'} $, averaged over $ \vec r' $ and
$ t' $ respectively. \\
We calculate the usual roughness and growth exponents (as defined in [1]) and compare these values with the analytical
prediction of the KPZ universality class, inspite of starting with a curvature dependent state. 

We studied the behavior of our 1D curvature dependent growth equation (eqn.(1)) using time steps of the order of $
\triangle t\:\sim\:0.001 $ unit and lower (upto $ \triangle t\:\sim\:{10}^{-5} $), with a system size of L = 1000. Later on 
we checked our results in a larger system (L = 10,000) too but the essential conclusions, to be shortly elaborated,
remained unchanged. The results were averaged over 1024 independent runs. We found that for small values of $
\lambda\:=\:0.2,\:0.3,\:0.35 $, the roughness and dynamic exponents converged exactly on the 1D KPZ values and the results are
shown in Figs. 1 and 2. 

\begin{figure}
\centerline{\psfig{figure=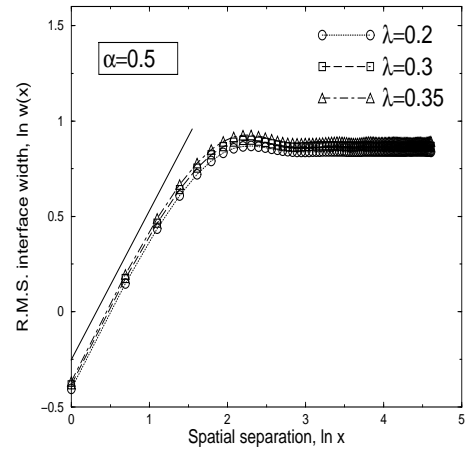,height=6.0cm,width=6.0cm,angle=-90}}
\caption{R.M.S. width against spatial separation in the log-log scale, with $ \lambda\:=\:0.2,\:0.3 $ and $ 0.35 $. This gives a measure of the roughness exponent $ \alpha (=0.5) $, which is seen to be exactly the KPZ value in 1D, as predicted from theory.}
\label{Fig:1}
\end{figure}

For the aforesaid values of $ \lambda $, the roughness exponents, ranged from 0.49 to 0.51 with a
fluctuation of $ \pm $0.02 


\begin{figure}
\centerline{\psfig{figure=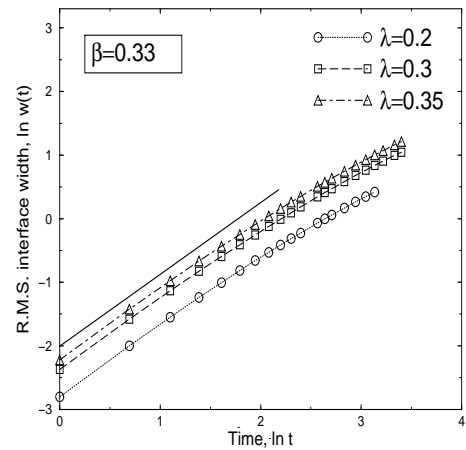,height=6.0cm,width=6.0cm,angle=-90}}
\caption{R.M.S. width against time in the log-log scale, with 
$ \lambda\:=\:0.2,\:0.3 $ and $ 0.35 $. This gives a measure of the growth exponent $ \beta (=0.33) $, which again gives the KPZ value in 1D, as per theoretical expectations.}
\label{Fig:2}
\end{figure}

and the dynamic exponents had the values between 0.334 to 0.337 with a 
fluctuation of $ \pm $
0.004 among them.
These values were obtained using V = 10 and a = 10 as the strength of the potential and lattice
spacing respectively. However, if the lattice constant is increased, which physically would suggest a more coarsened picture, keeping V
unchanged, the periodic potential has a much reduced contribution as
compared to the curvature term and in this case, there is a cross-over
to the Govind-Guo behavior [27], with $ \alpha\:=\:1.414\:\pm\:0.002 $ and $ \beta\:=\:0.392\:\pm\:0.01 $. Obviously,
this phase persists with all higher values of the lattice constant, keeping V unchanged. With larger values of V ($ \sim $ 100) or smaller
values of 'a'($ \sim $ 1 and smaller still), the periodic potential holds sway and muffles the other terms. Now the higher
harmonics of periodicity start popping up and this drastically modifies the overall scaling structure, such that the KPZ
scaling is totally lost. Hence, for all of our future discussions, we will restrict ourselves to the values V = 10 and a =
10 to avoid insurgence of the higher harmonics, which, in the process, gives a practical measure of the perturbative limit,
with respect to the parameters in hand. Although, it would have been most enterprising to have a theory in the
"strong-coupling limit" ({\it i.e.} large V, small 'a' and not a
field-theoretic definition) too, the fact that analytical techniques allow us only to trod into
the "perturbative" domain, we refrain from encroaching into the "forbidden" zones, even in the numerical estimates,
in the limit where the dynamics is dominated by strong periodic potentials. 

As stated already, a numerical analysis of the system, in the perturbative limit, clearly shows that for small enough values of $ \lambda $, the
model is in the isotropic KPZ universality class. This concurs with analytical predictions of the previous section too. But
in this section we deal with one more important property of the discretized system, that of evidences of "instability" in
the growth process, even in the perturbative limit. This is inaccessible through analytics and also by an atomistic
simulation process, since in both these cases, the possibility of instability is discarded by their definition themselves
[19]. At this point, it might seem reasonable to recapitulate the idea of "singularity" as observed in the earlier works
[19,32,33]."Singularity", in our case (as in the earlier references too), is defined as a sudden and a rapid growth of nearest neighbor height difference at
some arbitrary site, eventually leading to a computer overflow. Although this effect was initially observed by Tu [32] but
both he himself and Doherty, {\it et al} [33] wrongly attributed the
effect to "numerical artifacts". In any case, in 2D, they analyzed the
existence of a {\it strong-coupling} (field-theoretically) regime where the interface
develops a local divergence in finite time, but no such thing happened
in the conventional weak-coupling (in a field-theoretic sense) regime, our point of concern. Only later incisive forays
by Dasgupta, et al [19] and by Marsili and Bray [20] brought out the physics in the fore. As our observations also
indicate, the effect is a generic cause, characteristic to most discretized growth models, certainly in our model too. In connection, a very interesting result has only recently appeared in a work by Cuerno and Moro [34]. From a dynamic renormalization group study, they have found that in general, for all dimensions, in the overdamped sine-Gordon model, the linear surface diffusion mechanism appears only as an irrelevant perturbation to the basic growth dynamics, although the presence of the sine-Gordon damping might bring about cross-overs characterized by specific scaling properties. This is doubly satisfying in the context of our results, in the situation of unconserved dynamics, although without the presence of any curvature term. The
instability basically arises due to the sudden development of local hillocks or humps, depending on the direction of the
average velocity of growth. In our model, the determining parameters for this "instability" are the curvature coefficient
$ \lambda $ and the strength of the pinning potential V. Our investigations indicate that with a positive value of $
\lambda $, the instability associated is of the hillock type while for a negative $ \lambda $, the instability owes its
origin to the sudden development of a hump. Since we are looking in the "perturbative limit", as has been already stated, V
is kept fixed to a positive value. An idea as to the order of magnitude of the heights of these hills or humps is developed
from a study of the deterministic version of our model, which predicts an instability if $ \frac{3}{2}\lambda {h_0}^2 -
\frac{2\pi V}{a} \sin(\frac{2\pi h_0}{a}) - 10\gamma h_0\:>\:0 $, where $ h_0 $ is the height of the isolated hillock (or
groove). A non-trivial solution to the above transcendental-type inequality indicates an inverse dependence of $ \lambda $
and V to $ h_0 $, while $ h_0 $ increases as the lattice constant 'a' is increased. Since we are dealing with constant
values of V and 'a' (V = a = 10), this implies that as $ \lambda $ increases, $ h_0 $ decreases and certainly the
probability of developing an instability goes higher accordingly. Although, no such analytical prediction is possible for
the actual stochastic equation but this nature of $ \lambda $
dependence is still retained. Thus starting with an initial $ h_0 $ satisfying the previous inequality, we develop an instability
shortly and depending on the sign of $ \lambda $, the nature (hillock or hump) of this instability is then decided. It
should also be mentioned in connection, that for a flat surface, upto $ \lambda = 0.39 $, we did not find the existence of
any instability but for all values of $ \lambda $, starting 0.4 onwards, the generic instability was seen to develop.
Obviously, in making this assertion, we again tacitly assume fixed values of V and 'a'. Here one might argue from the above
result that if we reduce $ \lambda $ sufficiently, then even in the presence of initial hillocks or humps at t =
0, we can exclude the possibility of instability. But one has to keep in mind the comparative importance of the
curvature-dependent nonlinear term with the periodic potential and $ \lambda $ cannot be reduced arbitrarily. It is,
however, extremely significant that even with largest allowable values of $ \lambda $, the effect of curvature is washed out
and the system converges towards the KPZ fixed point in the perturbative limit. As a summary of this section, we can
convincingly argue that the instability observed is an inherent property of the discretized growth process itself and
cannot be removed by reducing the integration time step $ \triangle t $ (our check was upto the limit $ \triangle
t\:\sim\:{10}^{-5} $), neither is this a fallout of the discretization scheme employed. The latter claim is amply borne out
by the fact that even if we improve on a finite-difference scheme by the Runge-Kutta, the results still remain unchanged.
Before concluding this section, we should probably spare a word or two about the numerical results to expect in 2+1
dimensions. In this context, we have to unfortunately declare that due to the extremely sensitive dependence of $ \lambda $
and V on the instability criterion, the definition of a perturbative limit (as was V = a = 10 in d=1+1) appeared to be
rather confused. Thus we have no claims to any conclusive results in d = 2+1 and as such would avoid making any substantial
demands in this regard. Thus all the numerical results are strictly valid for d = 1+1, although, it might not be too wishful
to expect that the basic KPZ universality class is still retained in higher dimensions also, at least, so far as the
analytical RG calculations are to be believed.

\section{Controlling of instability}

As our previous discussions have already shown that we are dealing with a stochastic system which has inherent properties
to develop instabilities in specific parametric limits and thereby, destroy all possibilities of observing the statistical
properties of the system at large time scales, it is extremely necessary to have some method of controlling these
instabilities. In so doing, we resort to the resourceful references 
[19,21,35] dealing with noise reduction techniques to suppress instabilities
in the aforementioned (KPZ, Lai-Das Sarma, etc.) nonlinear discrete models. 
To control instability in our model, we use the "irrelevant (in the RG sense) 
operator method" as used in these references. 
The conventional way to tackle this is to introduce the "irrelevant variables" (in the RG sense) which in turn would
scale off to zero, compared to the leading nonlinearity in the dynamical equation. Thus the scaling behavior of the system
remains unchanged but in the process, these "irrelevant" operators kill the instability lurking within the discrete model.
For our case, we introduce an infinite series of irrelevant terms through the function f($ {|{\nabla}^2 h|}^2 $) on the
right hand side of eqn.(1), where

\bea
f(x) = \frac{1}{c}\:(1-exp(-cx))
\eea

{\noindent} 'c' being an adjustable parameter. For $ x << \frac{1}{c} $, f(x) approaches x and in the other extreme, for $
x >> \frac{1}{c} $, f(x) tends to a constant value 1/c. For our RC model, a value of c, even as small as 0.001 is
sufficient to stabilize the system from the instabilities. However, this is the critical limit for c, below which the
function f($ {|{\nabla}^2 h|}^2 $) loses its stabilizing property. Incidentally, this critical value of c in our model is
an order of magnitude lower than was obtained by Dasgupta, {\it et al} with more conventional KPZ or LD models. For any
further theoretical details on this topic, we refer the reader to the references [19,21] and avoid unnecessary repetition
of well-established information. As a graphical back-up of our claim to stabilization, we demonstrate Figs.3 and 4. 

\begin{figure}
\centerline{\psfig{figure=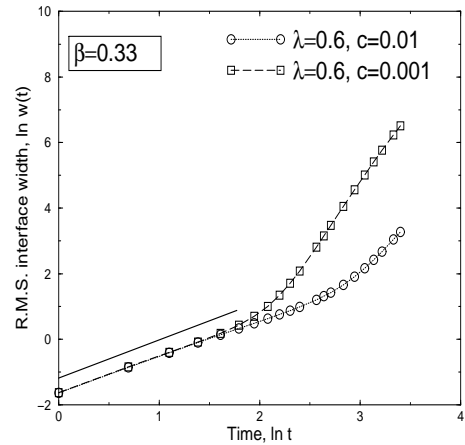,height=6.0cm,width=6.0cm,angle=-90}}
\caption{R.M.S. width against time in the log-log scale, with
$ \lambda\:=\:0.6 $ and $ c=0.01 $ and $ 0.001 $. This gives an idea of the efficacy of the instability controlled growth model.} 
\label{Fig:3}
\end{figure}

In Fig.3, 
we plot the r.m.s. width against time, for $ \lambda = 0.6, \:h_0=-10.0 $, using c values of 0.01 and 0.001. The linear
regions in the plots preserve the expected KPZ scaling with $ \beta = 0.337 $. In Fig.4, we plot the same quantities as in
Fig.3 but for $ \lambda = -0.6 $ and $ h_0 = -10.0 $ with identical c values. 

\begin{figure}
\centerline{\psfig{figure=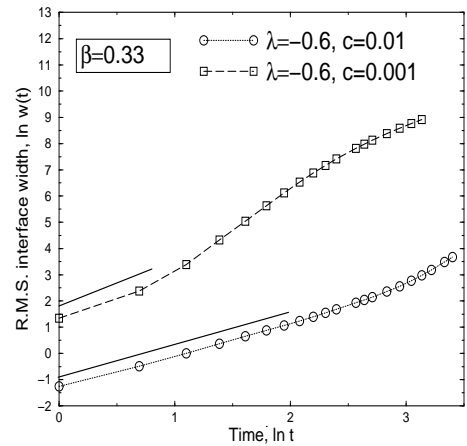,height=6.0cm,width=6.0cm,angle=-90}}
\caption{R.M.S. width against time in the log-log scale, with
$ \lambda\:=\:-0.6 $ and $ c=0.01 $ and $ 0.001 $. Thus with both $ \lambda $ and $ h_0 $ negative, due to the "hump"-type instability occuring here, the linear zone, towards the origin, is rather small with c=0.001.} 
\label{Fig:3}
\end{figure}

We see that again the linear regions in the
plots preserve the KPZ scaling and interestingly enough, now the system is stabilized even with both $ \lambda $ and $ h_0
$ negative, something which was not possible previously. However, as is quite evident from Fig.4, the process of stabilization is not automatic in the case where both the curvature-controlling parameter $ \lambda $ and $ h_0 $ are of the same sign. This happens due to the extremely strong unstable oscillations arising out of this combination and with c=0.001, only a very small linear zone (defining $ \beta=0.33 $) is achievable. Both in figures 3 and 4, departure from the linear behavior indicate the insurgence of instabilties leading to non-universal non-KPZ behaviors. Our numerical results with a larger system size, L=10,000 shows that the effect is even more pronounced with larger system sizes.   

We conclude this section with comments on the possibilities of multiscaling in our RC model. In all our discussions so
far, we have most deliberately refrained from making comments on this issue. This is because, due to the presence of the periodic
potential term in the dynamics, which, in effect, provides a pinning mechanism to the growth process, there are always
possibilities of appearance of higher harmonics for all sizable (meaning larger than the "perturbative" value) values of
the potential V. In fact, as we have stated even, for V larger than 10 units, the scaling behavior is lost altogether due
to the insurgence of these higher ordered periodicities. Now, naively, this is something to be expected too, since, our
analytical exercise strictly borders on a perturbative estimate which neglects possibilities of stronger coupling terms. As
soon as this assumption is vilified, the analytical comparisons no longer hold good and we land on soft grounds. In any
case, since multiscaling
exists only for a very small time, after which it is suppressed by instabilities, we do not expect to lose any important
dynamic information of the system in neglecting this particular study, in reference to our model.

\section{Discussions and conclusions}

Here we discuss and analyze results obtained from our analytical (in 2D) and numerical (in 1D) studies in the previous
sections. To begin with, an extensive dynamic renormalization group calculation of the restricted-curvature (RC) model in
2+1 dimensions proves that the system actually belongs to the isotropic KPZ universality class. This result has rather far
reaching consequences. The result implies that in our model of surface growth, which is naively expected to be explicitly controlled by 
the local curvature in the pinning-depinning limit, in actuality, the curvature dependence is overshadowed by the KPZ
nonlinearity arising out of the pinning potential, thereby leading to a KPZ fixed point in the long distance, large time
limit. Thus in a normal growth process, particles will in general have a greater tendency to roll-off to its nearest
neighbor sites, producing a lateral growth, than being controlled by the convex or concave structures of the locality,
possibly arising out of the local impurities. Hence we might venture to predict that the effect of impurities in a
dynamically growing surface would not show its effect by a curvature dependence of the locality. An identical conclusion,
though in a different situation, was arrived at by Marsili, et al [21] through a reparametrization argument of the basic
Langevin equation. There they convincingly argued that a contribution from a local metric, arising
out of a curved surface, does not have any effect on the overall growth process itself. 

The nonlinear exercise is largely a rethinking in the line of Dasgupta, et al [19], obviously in context to our curvature
dependent model. The first major conclusion which we arrive from a numerical integration of the continuum equation
(eqn.(1)) is that, at least, in 1+1 dimensions, the model belongs to the KPZ universality class. The numerics give clear
evidence of a suppression of possible effects of the local curvature and driving the system to a KPZ fixed point, exactly
as was expected from the more generalized 2+1 dimensional model, by analytical calculations. This means that there is an
automatic switch-over from the curvature dependent dynamics to the lateral growth dynamics, demanded by the system itself.
This fact has one other indirect back-up from numerical observations detailed in the later part of the manuscript, in that,
at arbitrary spatial sites, a sudden increase in the values of the nearest-neighbor height differences was seen to produce
instabilities and a resulting computer overflow. Our model also shows generic
instabilities, independent of the size of the integration time steps or the definition of the discretization scheme
employed. However, our model has two independent parameters defining the instability, which are $ \lambda $ and V, as opposed to only a
single one ($ \lambda $) in [19]. But in effect, since we are concerned only with the perturbative limit, the second
parameter in our case is a "mild" one, compared to $ \lambda $. An interesting observation in this regards is that, for
all negative V's, the system converges towards the Govind-Guo fixed point [27]. The reason, again, is not difficult to
understand, since the periodic potential, with a negative strength, only acts as a stabilizing higher-ordered perturbation
to the Govind-Guo phase and hence is rendered irrelevant in the long time limit. In the perturbative domain, the effect of
V ($ >\:0 $) shows up in reducing the value of $ \mathrm{c_{min}} $. As have already been stated, our numerical value of 
$ \mathrm{c_{min}} $ is an order of magnitude lower than that is the case with more conventional KPZ or LD type of
dynamics. 

To conclude, in discretized versions of nonlinear continuum models we encounter
generic instabilities in the sense that certain sites grow much faster
than their nearest-neighbor sites, although for linear growth models, the 
discrete and continuum theories tend to agree with each other. However, the 
numerical stabilization scheme
seems to be an extremely reliable one and one hopes that other works in the 
future will bear further testimony to the fact. Regarding
whether discretized versions of growth models are the more fundamental ones, as compared to the corresponding continuum 
versions, we believe there is an enormous scope of argument in this matter and our stance is that of a totally undecided
one. This is simply because although the discretized versions of a lot of continuum equations seem to throw up possibilities
of generic instabilities in systems which cannot be accessed from the coarsened dynamics, simultaneously we should not
forget that atomistic versions (by their definition themselves) do not
have any instability. To elucidate this further, we might consider
atomistic models like the restricted solid-on-solid model by Kim and
Kosterlitz [35] and the conserved version of that [36]. These models,
belonging to the KPZ [14] and the Lai-Das Sarma [15] universality
classes respectively do not show any instability, since the basic atomistic 
growth rules in these models do not allow any surface height fluctuations to
exceed a prefixed finite number, eg. unity in the Kim-Kosterlitz model. Thus by design, these models are taylor-made to suppress all such instabilities
that might arise in a growth process. 
What this means is that the coarse-grained
dynamical equations might have terms irrelevant in the RG sense
but these terms are certainly not present in their atomistic
analogues. The bottomline story that we get from this comparison of
the continuum models with atomistic models in the same universality
class is that any conclusions based on a "brute-force" integration of
the continuum equation is valid also for the atomistic case, {\it as
  long as there is no instability}. But in the unstable zone, where
normal scaling breaks down, such a comparison is not valid anymore. Thus the issue is far less
than settled and we would love to have evidences, bearing conclusive verdict on this issue, in the future. However, our
claim of the independence of a dynamic growth process to any explicit curvature effects, which is supported by both our
analytical and numerical observations (though in different dimensions), is something which can be tested from
simple-to-device experiments and we are eagerly looking forward to such experimental confirmation.

\section{Acknowledgement}

The author is thankful to J. K. Bhattacharjee, C. Dasgupta, B. K. Chakrabarti 
and A. Dutta for numerous illuminating discussions with them during the course 
of this work. 

\section{Appendix}

The complete renormalization flow equations of the restricted-curvtaure model,
upto the second order, are 

\bea
\frac{dU}{dl} = (4-n)U
\eea

\bea
\frac{d\gamma'}{dl} = \frac{2{\pi}^4}{\gamma a^4}\:n A^{(\gamma')}
(n;\kappa)\:U^2
\eea

\bea
\frac{d\gamma}{dl} = -\frac{{\pi}^4}{6\gamma a^4}\:n A^{(\gamma)}(n;\kappa)\:
U^2
\eea

\bea
\frac{d\eta}{dl} = \frac{8{\pi}^4}{\gamma a^4}\:\frac{\eta}{\gamma}\:n
A^{(\eta)}(n;\kappa)\:U^2
\eea

\bea
\frac{d\lambda}{dl} = 0
\eea

\bea
\frac{d\lambda'}{dl} = \frac{8{\pi}^5}{\gamma a^5}\:n A^{(\lambda')}
(n;\kappa)\:U^2
\eea

\bea
\frac{dK}{dl} = 4K + \frac{D}{4\pi \eta \gamma}\:\lambda - \frac{4{\pi}^3}
{\gamma a^3}\:n A^{(K)}(n;\kappa)\:U^2
\eea

\bea
\frac{dD}{dl} = \frac{1}{8\pi}\:D\:\frac{{\lambda}^2 D}{{\gamma}^3} +
\frac{8{\pi}^4}{\gamma a^4}\:\frac{D}{\gamma}\:n A^{(\eta)}(n;\kappa)\:U^2
\eea

where $ U = \frac{V}{{\Lambda}^2},\:\:K = \frac{F}{{\Lambda}^2},\:\:
\kappa = \frac{2\pi K}{a \gamma} $ and $ n = \frac{\pi T}{\gamma a^2} $, with
$ \gamma' $ representing the coefficient of the renormalized surface term
[31] and $ \lambda' $ denoting the coefficient corresponding to the KPZ
nonlinearity generated on account of renormalization. $ A^{(i)}(n;\kappa) $
stand as shorthand representation for the integrals, where $ i = \eta,
\lambda', \lambda, \kappa $, etc. and whose detailed functional forms are
given below:

\ber
A^{(\gamma')}(n;\kappa) &=& \int_0^{\infty}\:\frac{dx}{x}\:\int_0^{\infty}\:
d{\tilde \rho}\:{\tilde \rho}^3\:J_0(\tilde \rho) \times \cos[\frac{2\pi}{a}
\frac{K x {\tilde \rho}^2}{\gamma}] \nonumber \\
& & \times \int\:dp\:\cos[\frac{\tilde \rho}{\lambda} p]\:\:exp[- \frac{n}{\Lambda}
\frac{\eta}{\gamma} {\tilde \rho} x p^4] \nonumber \\
& & \times exp[-x{\tilde \rho}^2 - 2n
\phi(\tilde \rho,x)]
\eer

\ber
A^{(\eta)}(n;\kappa) &=& \int_0^{\infty}\:dx \int_0^{\infty}\:d{\tilde\rho}\:
{\tilde \rho}\:J_0(\tilde \rho) \times \cos[\frac{2\pi}{a} \frac{K x {\tilde
\rho}^2}{\gamma}] \nonumber \\
& & \times \int\:dp\:\cos[\frac{\tilde \rho}{\lambda} p]\:\:exp[- \frac{n}{\Lambda}
\frac{\eta}{\gamma} {\tilde \rho} x p^4)] \nonumber \\
& & \times exp[-x{\tilde \rho}^2 - 2n
\phi(\tilde \rho,x)]
\eer
 
\ber
A^{(\lambda')}(n;\kappa) &=& \int_0^{\infty}\:\frac{dx}{x}\:\int_0^{\infty}\:
d{\tilde \rho}\:{\tilde \rho}^3\:J_0(\tilde \rho) \times \sin[\frac{2\pi}{a}
\frac{K x {\tilde \rho}^2}{\gamma}] \nonumber \\
& & \times \int\:dp\:\cos[\frac{\tilde \rho}{\lambda} p]\:\:exp[- \frac{n}{\Lambda}
\frac{\eta}{\gamma} {\tilde \rho} x p^4)] \nonumber \\
& & \times exp[-x{\tilde \rho}^2 - 2n
\phi(\tilde \rho,x)]
\eer

\ber
A^{(K)}(n;\kappa) &=& \int_0^{\infty}\:\frac{dx}{x}\:\int_0^{\infty}\:
d{\tilde \rho}\:{\tilde \rho}\:J_0(\tilde \rho) \times \sin[\frac{2\pi}{a}
\frac{K x {\tilde \rho}^2}{\gamma}] \nonumber \\
& & \times \int\:dp\:\cos[\frac{\tilde \rho}{\lambda} p]\:\:exp[- \frac{n}{\Lambda}
\frac{\eta}{\gamma} {\tilde \rho} x p^4)] \nonumber \\
& & \times exp[-x{\tilde \rho}^2 - 2n
\phi(\tilde \rho,x)]
\eer 

\ber
A^{(\gamma)}(n;\kappa) &=& \int_0^{\infty}\:\frac{dx}{x}\:\int_0^{\infty}\:
d{\tilde \rho}\:{\tilde \rho}^5\:J_0(\tilde \rho) \times \cos[\frac{2\pi}{a}
\frac{K x {\tilde \rho}^2}{\gamma}] \nonumber \\
& & \times \int\:dp\:\cos[\frac{\tilde \rho}{\lambda} p]\:\:exp[- \frac{n}{\Lambda}
\frac{\eta}{\gamma} {\tilde \rho} x p^4)] \nonumber \\
& & \times exp[-x{\tilde \rho}^2 - 2n
\phi(\tilde \rho,x)]
\eer

{\noindent} Equations (8) to (15) contain the total dynamical response of the 
system around the kinetic roughening limit. The behaviors of 
these integrals, with variation in the values of $ \kappa $ are more or less 
identical to that of Fig.1 in [9], only in our case, the tails of each of these
 curves are more sharply defined, due to the additional multiplicative
 exponential factors appearing in each of our integrals. 

\end{multicols}
\end{document}